\documentstyle{article}

\newcommand{\mathrm}{\rm}

\begin{document}
\begin{center}
{\Large\bf Magnetic Monopoles and Massive Photons in a  Weyl-Type  
Electrodynamics.} 
\\[1cm]
{\bf
Mark Israelit$^{\dagger, }$\footnote{e-mail:
 Marc.Israelit@uni-konstanz.de , and:  israelit@physics.
technion.ac.il , (permanent e-address)}
}\\[0.5cm]
$^\dagger$ Department of Physics, University of Konstanz,
 PF 5560 M678, D-78434 Konstanz, Germany.

On leave from: Department of Physics, University of Haifa,
 Oranim, Tivon 36006, Israel.\\
\medskip

\end{center}

\begin{abstract}
\noindent
In a previous work the Weyl-Dirac framework was generalized
 in order to obtain a geometrically based general relativistic
theory, possessing intrinsic electric and magnetic currents and
 admitting massive photons.

Some physical phenomena in that framework are considered.
 So it is shown that massive photons may exist only in
presence of an intrinsic magnetic field. The role of massive
 photons is essential in order to get an interaction between
magnetic currents. A static spherically symmetric solution
is obtained. It may lead either to the Reissner-Nordstr\o m
metric, or to the metric created by a magnetic monopole.
\medskip


\end{abstract}

\vspace{0.5cm}

%%%%%%%%%%%%%%%%%%%%%%%%%%%%%%%%%%%%%%%%%%%%%%%%%%%%%%%%%%%%%%%%%%%%%
%%%%%%%%%%%%%%%%%%%%%%%%%%%%%%%%%%%%%%%%%%%%%%%%%%%%%%%%%%%%%%%%%%%%%

\section{Introduction}
Two fundamental electrodynamical phenomena are
standing beyond the frames of Maxwell's theory.
The first "outsider" being known as magnetic
monopole was evoked by Dirac \cite{Dirac1931},
\cite{Dirac1948}, (cf. also an interesting review
\cite{Blagojevic}), the second being the massive
photon . The massless photon became a tacit axiom
of physics own to the success of quantum
electrodynamics in predicting experiments with
enourmously high exactness. But the same results
 would be obtained with photons having mass
$m_{\gamma}<10^{-48}$ g. (cf.\cite{Goldhaber},
 \cite{Ignatiev}). From the quantum-theoretical
 standpoind Dirac's monopole and massive photons
  were discussed widely during the last decades
(\cite{Blagojevic}, \cite{Goldhaber}, \cite{Cabrera}).
 But a satisfactory classical framework including these
 two phenomena was absent untill the last time.
If one wants to discuss massive photons he has
to consider Proca's equations rather then the
Maxwell equations. Further, if a magnetic charge
(monopole) really exists, then  Maxwell's
electrodynamics that suffers from an asymmetry,
 regarding to electric, and   magnetic currents,
must be replaced by a generalized theory, with the
dual field tensor having a non-vanishing divergence.
It would be desirable to build up the framework
from geometrical reasons, starting from a generalization
 of Riemann's geometry. Recently a massive electrodynamics,
based on a space with non-metricity and torsion,
was proposed \cite{Israelit1996}. Electric, and intrinsic
magnetic currents, as well massive photons coexist
 within this framework. In the limiting case one
obtains the ordinary Einstein-Maxwell theory.

In the presemt work that theory is developed,
and some characteristic phenomena, and crucial
problems are considered. The interaction between
electric and magnetic currents  and fields,
as well the dynamical role of massive photons
are considered.
It is shown that in absence of fields, created by
magnetic charges no massive photons are allowed.
The energy-momentum conservation law is discussed,
and the equation of motion of a charged
(either magnetically, or electrically)  test particle
is derived from it. It is shown that two magnetic
monopoles interact by means of massive photons. A static,
 spherically symmetric solution for vacuum is obtained.
From it, by an appropriate choice of parameters,
one obtains two alternative solutions, either the
Reissner-Nordstr\o m  one for an electric monopole,
or the metric and magnetic field of a magnetic monopole.
Thus one can not have both, an electric, and a magnetic
charge located in one point. The magnetic monopole is
found to be a massive entity.

%XXXXXXXXXXXXXXXXXXXXXXXXXXXXXXXXXXXXXXXXXXXXXXXXXXXXXXXX%

\section{The Torsional Weyl-Dirac Electrodynamics}
Let us consider  in brief  the  Torsional Weyl-Dirac
Electrodynamics. Details   may be found in a previous
work of the present writer \cite{Israelit1996}.
We started from  Weyl's geometry  \cite{Weyl1919},
 as modified by Dirac \cite{Dirac1973}, and developed
 by Rosen \cite{Rosen1982}. It was assumed that there
are given a symmetric  metric tensor $g_{\mu\nu}$,
a Weyl connection vector $w_\mu$, the Dirac scalar
gauge function $\beta(x^{\mu})$, (as in the original
Weyl-Dirac theory \cite{Weyl1919}, \cite{Dirac1973},
\cite{Rosen1982}), and also a torsion tensor
$\Gamma^\lambda_{\,\left[\, \mu\nu\right]}$ in
each point of the 4-dimensional space-time manifold.
In this case the assymetric  connection
$\Gamma^\lambda_{\,\mu\,\nu}$ may be  written
as (cf. \cite{Schouten})
\begin{equation} \label{1}
\Gamma^\lambda_{\,\mu\,\nu}=\left\{^\lambda_{\mu\,\nu}
\right\}+g_{\mu\nu} w^\lambda - \delta^\lambda_\mu w_\nu -
\delta^\lambda_\nu w_\mu+C^\lambda_{\ \mu\,\nu}  ,
\end{equation}
where $\left\{^\lambda_{\mu\,\nu}\right\}$ is the
Chritoffel symbol formed with $g_{\mu\nu}$, and
the contorsion  is given  in terms of the torsion
tensor as follows
\begin{equation} \label{2}
 C^\lambda_{\,\mu\nu}=\Gamma^\lambda_{\,\left[\,  
\mu\nu\right]}-g^{\lambda\beta}g_{\sigma\mu}
\Gamma^\sigma_{\,\left[\beta\nu\right]}-
g^{\lambda\beta}g_{\sigma\nu}
\Gamma^\sigma_{\,\left[\beta\mu\right]}.
\end{equation}
The Weylian character of the connection
(\ref{1}) causes a non-integrability of
length, so that one is faced with  local
gauge transformations
\begin{equation}\label{3}
B\rightarrow\overline B=e^\lambda B\ ;\quad g_{\mu\nu}
\rightarrow\overline g_{\mu\nu}=e^{2\lambda}g_{\mu\nu} ;
\quad w_\mu\rightarrow\overline w_\mu=w_\mu+
\lambda_{, \mu };\quad \beta\rightarrow\overline\beta
 =e^{-\lambda}\beta.
\end{equation}
where $B^2=g_{\mu \nu}B^{\mu} B^{\nu}$ is the length
of a vector $B^{\mu}$, and  $\lambda$  is an arbitrary
function of the coordinates.

In this generalized torsional  Weyl geometry
it was  assumed  that the torsion tensor is gauge
invariant, so that in addition to (\ref{3}) one has
\begin{equation}\label{4}
\Gamma^\lambda_{\,\left[\,\mu\nu\right]} \rightarrow
\overline{\Gamma}^\lambda_{\,\left[\,\mu\nu\right]}=
\Gamma^\lambda_{\,\left[\,\mu\nu\right]} .
\end{equation}
The equations of the theory were derived from
a variational principle
\begin{equation}\label{5}
\delta I=0~~,
\end{equation}
with the  action  $I$ , formed from curvature
invariants of the torsional Weyl space
(cf.\cite{Israelit1996}).
\begin {eqnarray}\label{6}
I =\int \biggl(W^{\mu\nu}W_{\mu\nu}-\beta^2 R+
\beta^2(k-6)w_\mu w^\mu+ 2(k-6)\beta w^\mu
\beta_{,\mu}+k\beta_{,\mu} \beta_{,\underline\mu}   
+8\beta\Gamma^\alpha_{\,\left[\lambda\alpha\right]}
\beta_{,\underline \lambda}
\\
\nonumber
 +\beta^2(2\Gamma^\alpha_{\,\left[\mu\lambda\right]}
\Gamma^\lambda_{\,\left[{\underline\mu}\alpha\right]}
-\Gamma^\alpha_{\,\left[\sigma\alpha\right]}
\Gamma^\omega_{\,\left[{\underline\sigma}\omega\right]}  
+\Gamma^\alpha_{\,\left[\mu\lambda\right]}
\Gamma^\omega_{\,\left[\underline\mu\underline\lambda\right]}
g_{\alpha\omega}+
8\Gamma^\alpha_{\,\left[\sigma\alpha\right]}w^\sigma)\\
\nonumber
 +4 W_{\mu\nu;\alpha}\Gamma^\alpha_{\,\left[
\underline\mu\underline\nu\right]} +2\Lambda
\beta^4+ L_{matter} \biggr)(-g)^{1/2}\,dx
\end{eqnarray}.
In the action (\ref{6}) $R$ is the Riemannian
curvature scalar formed with the Christoffel symbols  
$\left\{^\lambda_{\mu\,\nu}\right\}$, the Weyl
curvature tensor is given by $W_{\mu \nu}=
w_{\mu , \nu}-w_{\nu , \mu}$, and $L_{matter}$
 is the Lagrangian density of matter.
Further, $k$ is an arbitrary parameter
(cf.\cite{Dirac1973}), $\Lambda$ is the
cosmological constant, an underlined index is
to be raised with the metric $g^{\mu\nu}$, a comma
stands for a  partial derivative, and a semicolom
(;) for a covariant derivative formed with  
$\left\{^\lambda_{\mu\,\nu}\right\}$.
The independent variables in (\ref{6}) are :
the metric tensor $g_{\mu\nu}$, the torsion tensor  
$\Gamma^\lambda_{\,\left[\,\mu\nu\right]}$
the Weyl connection vector $w_{\mu}$ and the
Dirac gauge function $\beta$.

In the original Weyl-Dirac theory (cf.\cite{Weyl1919},
\cite{Dirac1973}) the Weyl connection vector $w_{\mu}$
was treated as the  potential vector of the
electromagnetic field, while the Weyl curvature tensor
$W_{\mu\nu}$ yielded the field tensor.
Here,  the divergence of the  torsion enters into
the field tensor, so that one has a dual field
with  non-vanishing divergence, and hence an
intrinsic magnetic current is present.

From the variational principle (\ref{5}), (\ref{6})
one obtains the following equation for the
electromagnetic field,
\begin {equation}\label{7}
\Phi^{\mu\nu}_{\, ; \nu}
=(1/2)(k-6)\beta^2 W^\mu+4 \pi J^\mu \quad,
\end {equation}
where the electromagnetic  field is introduced as
\begin {equation}\label{8}
\Phi_{\mu\nu}=W_{\mu\nu}-
2\Gamma^\alpha_{\,\left[\mu\nu\right];\alpha}\equiv
W_{\mu;\nu}-W_{\nu;\mu}-
2\Gamma^\alpha_{\,\left[\mu\nu\right];\alpha}\quad.
\end{equation}
and the dual is  defined in the usual manner
\begin {equation}\label{9}
\widetilde{\Phi}^{\mu\nu} =-\frac{1}{2  
(-g)^{1/2}}\;\varepsilon^{\mu\nu\alpha\beta}
\Phi_{\alpha\beta} \quad ,
\end{equation}
with $\varepsilon^{\mu\nu\alpha\beta}$
standing for the completely antisymmetric
Levi-Civita symbol, and $\varepsilon^{0123}=1$.

For the dual field one obtains
\begin {equation}\label{10}
\widetilde\Phi^{\mu\nu}_{\, ; \nu}=-2\pi L^{\mu} \quad.
\end{equation}
In equations (\ref{7}), and  (\ref{10}) the followig
quantities are introduced: $W_{\mu}$ stands for the
gauge invariant Weyl vector
\begin {equation}\label{11}
W_\mu=w_\mu+(\ln\beta)_{,\,\mu}\quad ,
\end{equation}
the electric current density is given by
\begin {equation}\label{12}
16\pi J^{\mu}=\frac{\delta L_{matter}}
{\delta w_{\mu}}\quad ,
\end{equation}
and the magnetic current density vector $L_{\mu}$
is introduced by
\begin {equation}\label{13}
L^{\sigma}=-\frac{1}{6 (-g)^{1/2}}\;
\varepsilon^{\sigma\mu\nu\lambda}
(\Omega_{\mu\left[\nu\lambda\right]}+
\Omega_{\lambda\left[\mu\nu\right]}+
\Omega_{\nu\left[\lambda\mu\right]})
\quad ,
\end{equation}
with a $\Omega_{\lambda\left[\mu\nu\right]}
\equiv g_{\mu\alpha}g_{\nu\beta}
\Omega^{\,\left[\alpha\beta\right]}_{\lambda}$,
 where the quantities $\Omega^{\,\left[\alpha
\beta\right]}_{\lambda}$ are defined  as

\begin {equation}\label{14}
16\pi\Omega^{\,\left[\mu\
nu\right]}_{\lambda}=\frac{\delta
L_{matter}}{\delta
\Gamma^\lambda_{\,\left[\mu\nu\right]}} \quad .
\end{equation}
From the field equations (\ref{7}),
(\ref{10}) one has the following current
conservation laws:
\begin {equation}\label{15}
(k-6)(\beta^2 W^{\mu})_{\, ;\mu} + 8\pi
J^{\mu}_{\, ; \mu}=0 ,
\end{equation}
and
\begin {equation}\label{16}
L^{\mu}_{\; ;\mu}=0 \quad ,
\end{equation}

Generally the torsion can be broken into
three irreducible parts
(cf. e.g.  \cite{Hammond} , \cite{Hayashi}):
a trace part, a traceless one, and a totally
antisymmetric part. It turns out that only the
third part is relevant  in our case, so that
the totally antisymmetric torsion tensor
$\Gamma^{\lambda}_{\left[\mu\nu\right]} $ may
be represented by a vector. If we
introduce the auxiliary torsion tensors:
\begin {equation}\label{17}
\Gamma_{\lambda\left[\mu\nu\right]}=g_{\sigma\lambda}
\Gamma^{\sigma}_{\left[\mu\nu\right]} ; \qquad
\Gamma^{\lambda\left[\mu\nu\right]}=
g^{\alpha\mu}g^{\beta\nu}
\Gamma^{\lambda}_{\left[\alpha\beta\right]} \quad .
\end{equation}
we can express the torsion  by means of a
gauge invariant vector $ V_{\mu}$
(named below torsion vector) as follows
\begin {equation}\label{18}
\Gamma_{\lambda\left[\mu\nu\right]}=(-g)^{1/2}\;  
\varepsilon_{\lambda\mu\nu\sigma}V^{\sigma} ;
\qquad
\Gamma^{\lambda\left[\mu\nu\right]}=-(-g)^{-1/2}\;  
\varepsilon^{\lambda\mu\nu\sigma}V_{\sigma}
\quad.
\end{equation}
This leads to
\begin {equation}\label{19}
\Gamma^{\nu}_{\left[\mu\nu\right]}=0  .
\end{equation}
From (\ref{18}) one can also derive the
following auxiliary formulae:
\begin {equation}\label{20}
\Gamma^{\lambda\left[\mu\nu\right]}_{\;\; ;
\lambda}=
\frac{\varepsilon^{\mu\nu\alpha\sigma}}
{2(-g)^{1/2}}(V_{\alpha\,; \sigma}-
V_{\sigma \,;\alpha}) \quad ;\quad \quad  
\Gamma^{\lambda\left[\mu\nu\right]}_{\;\; ;
\lambda \, ; \nu}=0~.
\end{equation}

Further by the choice(\ref{18}) one has
from (\ref{8})

\begin {equation}\label{21}
\Phi^{\mu\nu}=(W^\mu_{;\,\underline\nu}-
W^\nu_{;\,\underline\mu})-
\frac{\varepsilon^{\mu\nu\alpha\sigma}}{(-g)^{1/2}}
(V_{\alpha\,;\,\sigma}-V_{\sigma\,;\,\alpha}) \quad ,
\end {equation}
and for the dual field (\ref{9})
\begin {equation}\label{22}
\widetilde\Phi^{\mu\nu}=-2(V^\mu_{;\,\underline\nu}-
V^\nu_{;\,\underline\mu})-
\frac{\varepsilon^{\mu\nu\alpha\sigma}}{2(-g)^{1/2}}
(W_{\alpha\,;\,\sigma}-
W_{\sigma\,;\,\alpha}) \quad.
\end {equation}

Inserting (\ref{21}) into  the field equation
(\ref{8}), and making use of (\ref{20}), we obtain
\begin {equation}\label{23}
\Phi^{\mu\nu}_{\, ; \nu}=   
W^\mu_{;\underline{\nu};\nu}-W^\nu_{;\underline{\mu};
\nu}
=(1/2)(k-6)\beta^2 W^\mu+4 \pi J^\mu \quad,
\end {equation}
while for the dual field we obtain from (\ref{10}),
 and (\ref{22})

\begin {equation}\label{24}
\widetilde\Phi^{\mu\nu}_{\, ; \nu}=(V^{\mu}_{;\,
\underline{\nu} ;\,\nu} -V^{\nu}_{;\,\underline{\mu};
\nu})
=-2\pi L^{\mu}
\quad.
\end{equation}
From (\ref{23}) one sees that
\begin {equation}\label{25}
\Phi^{\mu\nu}_{\, ;\,\nu}=W^{\mu\nu}_{\,; \,\nu}  .
\end {equation}

Varying in (\ref{6}) the metric tensor $g_{\mu\nu}$,
one obtains the equation for the gravitational field
\begin{eqnarray}\label{26}
G^{\mu\nu}= -(8\pi/{\beta^2})  
T^{\mu\nu}-(8\pi/{\beta^2})(\widetilde{M}^{\mu\nu}-
\overline {M}^{\mu\nu})+
(2/\beta)(g^{\mu\nu} \beta_{\, ;\underline{\alpha}
\,;\alpha}-
\beta_{\, ;\underline{\mu}\, ;\underline{\nu}})
\\
\nonumber
+(1/\beta^2)(4\beta_{; \underline{\mu}}\beta_{;
\underline{\nu}}-g^{\mu\nu}
\beta_{; \underline{\alpha}} \beta_{;\alpha})+
(k-6)(W^\mu W^\nu -\frac{1}{2} g^{\mu\nu}W^\sigma
W_\sigma) -g^{\mu\nu} V^\sigma V_\sigma-2V^\mu V^\nu .
\end{eqnarray}
 where $8\pi T^{\mu\nu}=\delta L_{matter}/
\delta g_{\mu\nu} $, and the modified energy density
tensors of the field are defined as follows
\begin {equation}\label{27}
4\pi\widetilde{M}^{\mu\nu}=(1/4)g^{\mu\nu}
\Phi^{\alpha\beta} \Phi_{\alpha\beta} -
\Phi^{\mu\alpha}\Phi^\nu_{\, \alpha} \quad ,
\end{equation}
and
\begin {equation}\label{28}
4\pi\overline{M}^{\mu\nu}=(1/4)g^{\mu\nu}
(\Phi^{\alpha\beta}-W^{\alpha\beta})
(\Phi_{\alpha\beta}-W_{\alpha\beta} ) -
(\Phi^{\mu\alpha}-W^{\mu\alpha})
(\Phi^\nu_{\,\alpha}-W^\nu_{\,\alpha}) \quad .
\end{equation}

It is remarkable (cf. (\ref{23})) that the
Weyl vector $W_{\mu}$ is created  either by
the electric currents  $J^\mu$, or by a Proca-type
self-inducing term. On the other hand
(cf.(\ref{24})), the torsion vector $V_{\mu}$
is created by the magnetic current density vector
$ L^\mu $.

\section{The Einstein Gauge}
The torsionless Weyl-Dirac theory with $k=6$
turns into the Einstein-Maxwell theory if one
chooses the Einstein gauge $\beta=1$. (cf.
\cite{Dirac1973},\cite{Rosen1982},
\cite{Isr.Rosen1983}).

Here  we consider in that gauge the above
treated generalized Weyl-Dirac theory, possessing
torsion, and allowing arbitrary values for $k$.
Turning to the Einstein gauge, we set
\begin {equation}\label{29}
\beta=1 .
\end{equation}
Making use of (\ref{29}), and replacing $W_{\mu}$
 by $w_{\mu}$  (cf.(\ref{11})), we obtain from
(\ref{26})
\begin{eqnarray}\label{30}
G^{\mu\nu}=-8\pi T^{\mu\nu}-8\pi
(\widetilde{M}^{\mu\nu}-
\overline{M}^{\mu\nu})
-(k-6)(w^{\mu} w^{\nu}-(1/2)g^{\mu\nu}
w^{\sigma} w_{\sigma})
\\
\nonumber
 -2V^{\mu} V^{\nu}- g^{\mu\nu} V^{\sigma}
V_{\sigma} \quad .
\end {eqnarray}
The fields in the Einstein gauge may
be written as
\begin {equation}\label{31}
\Phi^{\mu\nu}=(w^\mu_{;\,\underline\nu}-
w^\nu_{;\,\underline\mu})-
\frac{\varepsilon^{\mu\nu\alpha\sigma}}{(-g)^{1/2}}
(V_{\alpha\,;\,\sigma}-V_{\sigma\,;\,\alpha}) \quad ,
\end {equation}
and
\begin {equation}\label{32}
W_{\mu\nu}=w_{\mu \,;\,\nu}-w_{\nu \,;\,\mu}~.
\end{equation}
The energy conservation  law  can  be obtained
from (\ref{30})  making use of the conctracted
Bianchi identities, so that
%%We have  $ G^{\nu}_{\mu ; \nu}=0 $ , so that
(\ref{60}) leads to
\begin{equation}\label{33}
8\pi ( T^{\nu}_{\mu ;\,\nu}+
\widetilde{M}^{\nu}_{\mu ; \,\nu} -
\overline{M}^{\nu}_{\mu ;\,\nu})+(k-6)
\left(w_{\mu} w^{\nu}-(1/2)\delta^{\nu}_{\mu}
w_{\sigma} w^{\sigma}\right)_{;\, \nu} +2(V_{\mu}
V^{\nu}) _{;\,\nu}+ (V_{\sigma} V^{\sigma})_{;\,\mu}
 =0 .
\end{equation}
Making in (\ref{33}) use of definitions
(\ref{27}), and (\ref{28}), and of equations
(\ref{23}), (\ref{24}), (\ref{25}),  one obtains
\begin{eqnarray}\label{34}
8\pi( T^{\nu}_{\mu \, ; \,\nu}+\Phi_{\mu\sigma}
 J^{\sigma}) +4\pi \sqrt{-g}\,\,
\varepsilon_{\alpha\beta\mu\sigma} W^{\alpha\beta}
 L^{\sigma}+(k-6)
(\Phi_{\mu\sigma}+W_{\mu\sigma})w^{\sigma}
\\
\nonumber
+(k-6) w_{\mu} w^{\nu}_{; \,\nu} +   2V^{\nu}
(V_{\mu \, ; \, \nu}+V_{\nu \, ; \, \mu}) +
2V_{\mu} V^{\nu}_{\, ; \,\nu} =0 \quad.
\end {eqnarray}
For a moment let us go back to the field
equations (\ref{23}), and (\ref{24}).
Equation (\ref{24}) remains unchanged,
while (\ref{23}), with   a new parameter
$\kappa^2 \equiv (1/2)(6-k) $, takes on the form

\begin {equation}\label{35}
\Phi^{\mu\nu}_{\, ; \nu}=  w^\mu_{;
\underline{\nu};\nu}-w^\nu_{;
\underline{\mu};\nu}
=-\kappa^{2} w^\mu+4 \pi J^\mu \quad.
\end {equation}
In absent of electric currents in a
certain region we obtain from (\ref{15}),
and (\ref{29})
\begin{equation}\label{36}
w^{\nu}_{\, ;\,\nu}=0 \quad,
\end {equation}
so that equation (\ref{35}) may be rewritten
in the following form
\begin{equation}\label{37}
w^{\mu}_{ ;\,\underline{\nu};\,\nu} +w^\nu
R_{\nu}^{\mu}+\kappa^2w^\mu=0   \quad ;
\end {equation}
with $R_{\nu}^{\mu}$ being the Ricci tensor,
formed from the usual Christoffel symbols. If the
curvature in the current-free region  is negligible,
one obtains  the Proca \cite{Proca1936} equation for
the vector field $w^{\mu}$
\begin {equation}\label{38}
w^{\mu}_{ ;\,\underline{\nu};\,\nu}+
\kappa^2  w^\mu = 0.
\end {equation}

From the quantum mechanical standpoint this
equation describes a particle having spin $1$ and
mass that in  conventional units is given by
\begin {equation}\label{39}
m_{\gamma}=(\hbar/c)\kappa=(\hbar/c)
\sqrt{\frac{6-k}{2}} ,
\end {equation}
thus, for $k<6$ one obtains massive field
particles, photons.

In the special case when, $V^{\mu}=0$,
and $k=6$, one obtains  from (\ref{24})
$L^{\mu}=0$, so that equations (\ref{30}),
and (\ref{35})   turn into the   equations
of  the Einstein-Maxwell theory, while  (\ref{34})
 becomes the  usual energy conservation law.

Let us go back to the conservation law (\ref{34}),
and consider the case of vacuum, so that

\begin {equation}\label{40}
T^{\nu}_{\mu}=0~~;~~~J^{\sigma}=0~~;~~~L^{\sigma}=0~~.
\end {equation}
in addition in vacuum condition (\ref{36})
is holding, and hence one is left with
\begin {equation}\label{41}
-2\kappa^2 (\Phi_{\mu\sigma}+W_{\mu\sigma})w^{\sigma}
 +   2V^{\nu} (V_{\mu \, ; \, \nu}+
V_{\nu \, ; \, \mu}) +
2V_{\mu} V^{\nu}_{\, ; \,\nu} =0 \quad.
\end {equation}
One readily sees from (\ref{41}) that the
condition  $V_{\mu}=0$, leads to ~$\kappa=0$~,
and hence in absence of magnetic fields massive
photons do not exist, so that the classical
Maxwell electromagnetism has only massless photons.    .

From the energy-momentum conservation law (\ref{34})
taking into account (\ref{36}), and (\ref{41}),
we can obtain the equation of motion of a test
particle, having mass (rest energy) $m_0$  electric
charge $\varepsilon_0$. and four-velocity $u_{\mu}$
in a given external field
\begin{equation} \label{42}
\frac{du^{\mu}}{ds}+
\left\{^\mu_{\lambda\,\nu}\right\}u^{\lambda}u^{\nu}
+(\varepsilon_0/m_0) u_{\nu}\Phi^{\mu\nu} =0 \quad.
\end {equation}
For a test particle having a magnetic charge
$\mu_0$ instead of an electric one we obtain
\begin{equation} \label{43}
\frac{du^{\mu}}{ds}+
\left\{^\mu_{\lambda\,\nu}\right\}u^{\lambda}
u^{\nu}+(1/2)(\mu_0/m_0) u^{\sigma}
\sqrt{-g}\,\,
\varepsilon_{\alpha\beta\lambda\sigma}
g^{\lambda\mu} W^{\alpha\beta}
 =0 \quad.
\end {equation}

One can imagine a charged test particle
moving in the neighbourhood of electrically,
and magnetically charged massive bodies.
According to (\ref{24}), (\ref{25}), and (\ref{35})
the torsion vector $V_{\mu}$ is created by magnetic
charged bodies, while the Weyl vector $w_{\mu}$ may
be created  by bodies having electric charge, as well
by a self-inducing Proca term. Now, the structure of
the field strength tensors is given by (\ref{31}),
and (\ref{32}). Thus, according to (\ref{42}),
on an electrically charged  test particle act both,
electric, as well magnetic sources. These interactions
are possible with both, massive or massless photons.
On the other hand, according to  (\ref{43}), a magnetically
 charged test particle is accelerated by the field
$W^{\mu\nu}$, so that it is affected by electrically
charged bodies, and by massive photons. One might
claim that there is no interaction between  magnetic
monopoles. However from (\ref{41}) one sees that massive
photons may accompany the magnetic torsion vector field.
Thus, two magnetic monopoles interact by means of massive
 photons.
The field of massive photons invoked by a magnetic
monopole is considered in section 6.

\section{Spherical Symmetry }

Suposse there is a particle at rest in the origin,
and we consider a region around the particle which
is so small that we can neglect the cosmic curvature.
 The static spherically symmetric line-element may
be written as
\begin{equation} \label{44}
ds^2=e^{\nu}dt^2-e^{\lambda}dr^2-r^2(d \vartheta^2+
\sin^2 \vartheta d \varphi^2)~,
\end {equation}
with $\lambda$, and $\nu$ being functions of $r$.
From symmetry reasons one can prove that there is
only one non-vanishing component of the Weyl vector
 $w_{\mu}$
\begin{equation} \label{45}
 w_0 \equiv w(r)~,
\end {equation}
and also one non-vanishing component of the
magnetic vector $V_{\mu}$
\begin{equation} \label{46}
 V_0 \equiv V(r)~.
\end {equation}
Let us write the Einstein equations (\ref{30})
in brief as:

\begin{equation} \label{47}
 G^{~\nu}_{\mu}=-8\pi E^{~\nu}_{\mu}.
\end {equation}
Making use of (\ref{22}), (\ref{27}), (\ref{28}),
(\ref{31}), as well as of (\ref{29}) , we obtain
for the metric (\ref{44}) the following non-zero
components of $E^{~\nu}_{\mu}$
\begin{equation} \label{48}
8\pi E^{~0}_0=8\pi T^{0}_0+ e^{-(\lambda+\nu)}(w')^2-
\kappa^2 e^{-\nu} w^2 +3~e^{-\nu}~ V^2~;
\end{equation}
\begin{equation} \label{49}
8\pi E^{~1}_1=8\pi T^{1}_1 +e^{-(\lambda+\nu)}(w')^2+
\kappa^2 e^{-\nu} w^2 +e^{-\nu}~ V^2~;
\end{equation}
and
\begin{equation} \label{50}
8\pi E^{~2}_2=8\pi E^{~3}_3=8\pi T^{2}_2-e^{-(\lambda+
\nu)}(w')^2+\kappa^2 e^{-\nu} w^2 +e^{-\nu}~ V^2~.
\end{equation}
with $ f'\equiv df/dr $.
 If we substitute  (\ref{48}) - (\ref{50}) into
(\ref{47}) we obtain the Einstein equations explicetly
\begin{equation} \label{51}
e^{-\lambda}\left(-\frac{\lambda'}{r}+\frac{1}{r^2}
\right)-\frac{1}{r^2} =-8\pi T^{0}_0-e^{-(\lambda+\nu)}
(w')^2+\kappa^2 e^{-\nu} w^2 -3~e^{-\nu}~ V^2~;
\end{equation}
\begin{equation} \label{52}
e^{-\lambda}\left(\frac{\nu'}{r}+\frac{1}{r^2}
\right)-\frac{1}{r^2}=-8\pi T^{1}_1-
e^{-(\lambda+\nu)}(w')^2-\kappa^2
e^{-\nu} w^2 -e^{-\nu}~ V^2~;
\end{equation}
and
\begin{equation} \label{53}
e^{-\lambda}\left[\nu''+\frac{1}{2}(\nu')^2+
\frac{1}{r}(\nu'-\lambda')-
\frac{1}{2}\lambda'\nu' \right]=-8\pi T^{2}_2  
+2e^{-(\lambda+\nu)}(w')^2-2\kappa^2 e^{-\nu}
 w^2 -2e^{-\nu}~ V^2~.
\end{equation}

Taking into account (\ref{44}), (\ref{45}),
and (\ref{46}) one can rewrite
 (\ref{35}), and (\ref{24}) accordingly as follows
\begin{equation} \label{54}
w''-\frac{1}{2}(\lambda'+\nu')w'+\frac{2}{r}w'=
\kappa^2 e^{\lambda}w-4\pi e^{\lambda}J_0~;
\end{equation}
and
\begin{equation} \label{55}
V''-\frac{1}{2}(\lambda'+\nu')V'+\frac{2}{r}V'=
2\pi e^{\lambda} L_0~.
\end{equation}
Integrating (\ref{54}) one writes
\begin{equation} \label{56}
w'=\frac{e^{1/2(\lambda+\nu)}}{r^2}\left\{q(r) +
\kappa^2 I(r)+Q\right\}~,
\end{equation}
where the electric charge within a sphere of radius
 $r$ is given by
\begin{equation} \label{57}
q(r)=-4\pi \int_{0}^{r} J_0 e^{1/2(\lambda-\nu)}
r^2 dr = 4\pi \int_{0}^{r} \rho_e e^{\lambda/2}
 r^2 dr~,
\end{equation}
the Proca "charge"  is given by
\begin{equation} \label{58}
I(r)=\int_{0}^{r} e^{1/2(\lambda-\nu)} w r^2 dr~,
\end{equation}
and $Q=\mathrm{const}$ is the charge located in
the origin.
Similarly we obtain from (\ref{55})
\begin{equation} \label{59}
V'=\frac{ e^{1/2(\lambda+\nu)}}{r^2 }\left\{l(r)+
\widetilde M \right\}~,
\end{equation}
where $\widetilde M=\mathrm{const}$~~ stands for
the magnetic charge located in the origin, and
\begin{equation} \label{60}
l(r)=2\pi \int_{0}^{r} e^{1/2(\lambda-\nu)} L_0 r^2 dr~.
\end{equation}

In addition, making use of (\ref{56}), we can
rewrite the energy condition (\ref{34}), stemming
from the Bianchi identity, as follows

\begin{equation} \label{61}
4\pi~[ (T^{~1}_1) '+\frac{1}{2} \nu  
'(T^{~1}_1-T^0_0)+\frac{~2}{r}(T^{~1}_1-T^{~2}_2) -
J_0 w' e^{-\nu}] +[2~\kappa^2 w w'+V V'-V^2 \nu'] ~
e^{-\nu}=0~.
\end{equation}

We can consider (\ref{56}) and (\ref{59}) together
with the system of equations (\ref{51}) - (\ref{53}).
Alternatively we can make use by (\ref{61}), instead
 of (\ref{53}).

\section{A Simple Vacuum  Solution}

One can think about vacuum  surrounding
the particle, and about massless photons,
so that.
\begin{equation} \label{62}
T^{\nu}_{\mu}=0~~;~~~J^{\sigma}=0~~;~~~
L^{\sigma}=0~~;~~~\kappa=0~~;
\end{equation}
and one is left with the following expressions
for the fields (cf. (\ref{56}), (\ref{59}))
\begin{equation} \label{63}
w'=\frac{e^{1/2(\lambda+\nu)}}{r^2} Q~~~,
\end{equation}
and
\begin{equation} \label{64}
V'=\frac{ e^{1/2(\lambda+\nu)}}{r^2 }
\widetilde M ~~.
\end{equation}
Further, making use of (\ref{62}), and of
(\ref{63}), one has from (\ref{51})
\begin{equation} \label{65}
e^{-\lambda}\left(-\frac{\lambda'}{r}+
\frac{1}{r^2}\right)-\frac{1}{r^2} =-
\frac{Q^2}{r^4}-3~e^{-\nu}~ V^2~;
\end{equation}
On the other hand from  (\ref{51}),
and (\ref{52}) one obtains
\begin{equation} \label{66}
e^{-\lambda}[\frac{\lambda ' + \nu '}{r}]=
2e^{-\nu}~ V^2 ~~,
\end{equation}
and  by (\ref{62}) one has from (\ref{61})
\begin{equation} \label{67}
V V'-V^2 \nu'=0~.
\end{equation}
Thus one can write
\begin{equation} \label{68}
V= K e^{\nu};~~ (K=\mathrm{const}).
\end{equation}
Inserting  (\ref{68}) into  (\ref{66}), and
integrating we obtain
\begin{equation} \label{69}
e^{\lambda+\nu}=\frac{1}{B^2 -K^2 r^2}~~,
\end{equation}
with $B$ being an arbitrary constant.

Let us go back to eq.(\ref{65}). With an auxiliary
 function
\begin{equation} \label{70}
 y=e^{-\lambda}~,
\end{equation}
and by (\ref{69}) we can rewrite it as
\begin{equation} \label{71}
\frac{1}{r} y ~' +\frac{1}{r^2} y + \frac{3K^2 y}
{B^2-K^2 r^2}=\frac{1}{r^2}-\frac{ Q^2}{r^4}~~.
\end{equation}
The solution may be written as
\begin{equation} \label{72}
y \equiv e^{-\lambda}=\frac{1}{B}
 \left( \frac{1}{r^2}-
\frac{K^2}{B^2} \right)
\left[B r^2+Q^2(B^2-2K^2r^2)\right]+ \frac{K_{1}
(B^2-K^2 r^2)^{3/2}}{B^3 r}~~;
\end{equation}
and, making use of (\ref{69}), we get
\begin{equation} \label{73}
e^{\nu}= \frac{1}{B^3 r^2}\left[B r^2 +
Q^2(B^2-2K^2 r^2)\right]+\frac{ K_{1}
(B^2-K^2 r^2) ^{1/2}}{B^3 r}~,
\end{equation}
where $K_1$ is a constant. Let us take
 $K_1=-2m$ with $m$ being the  particle mass.
 The two other constants, $Q$   and $K$ represent
the electric charge and the magnetic field
(cf. (\ref{63}), and (\ref{68})) respectively.
In order to get for vanishing $Q$, and $K$ from
(\ref{72}), (\ref{73}) the Schwarzschild
metric we must set
\begin{equation} \label{74}
B=1~.
\end{equation}
If magnetic monopoles are absent, we can set
$K=0$~, and take into account (\ref{74}), so
that (\ref{72}), (\ref{73}) yield the Reissner-Nordstr\o m
metric:
\begin{equation} \label{75}
e^{\nu}=e^{-\lambda}=1+\frac{Q^2}{r^2} -
\frac{2m}{r}~~,
\end{equation}

Now, let us go back to (\ref{64}), and
(\ref{68}). Taking into account (\ref{74})
 we have from (\ref{68}):
\begin{equation} \label{76}
V'=\varepsilon K \left(e^{\nu}\right)'=
\varepsilon K \left[- \frac{2Q^2}{r^3}+
\frac{2m}{r^2 (1-K^2 r^2)^{1/2}} \right]~.
\end{equation}
Where the unit charge (electric, as well as
magnetic) $\varepsilon$ is introduced in order
to keep the dimensions identical with those of
general relativity. From (\ref{76}), and (\ref{69})
 one readily sees that (\ref{64})
is satisfied only for $Q=0$. In this case we have
 from (\ref{72}), and (\ref{73})
\begin{equation} \label{77}
e^{-\lambda}=\left(1-K^2 r^2\right) \left[1 -
\frac{2m}{r}(1-K^2 r^2)^{1/2}\right]~,
\end{equation}
and
\begin{equation} \label{78}
e^{\nu}=1-\frac{2m}{r}(1-K^2 r^2)^{1/2}~;
\end{equation}
while for the magnetic field strength we have
from (\ref{76})
\begin{equation} \label{79}
\widetilde\Phi_{01}=V'=\varepsilon K \left(e^{\nu}
\right)'=\frac{2m \varepsilon K}{r^2
(1-K^2 r^2)^{1/2}}~.
\end{equation}
Comparing this with (\ref{64}), and making
use of (\ref{69}) we obtain
\begin{equation} \label{80}
K= \frac{\widetilde M}{2 \varepsilon m}~.
\end{equation}
The solution (\ref{77}) - (\ref{79}) is defined for
\begin{equation} \label{81}
0<r<r_b=\frac{1}{K}=\frac{2 \varepsilon m }
{\widetilde M}
\end{equation}
There exists also such a radius
\begin{equation} \label{82}
r=r_s=\frac{2m}{(1+\widetilde M^2/
\varepsilon^2)^{1/2}}~,
\end{equation}
that from (\ref{77}), (\ref{78}) one has
\begin{equation} \label{83}
e^{\nu}(r_s)=e^{-\lambda}(r_s)=0~.
\end{equation}
Tus the particle is surrounded by a surface, on
which the metric is singular. For $\widetilde M =0$,
it turns into the Schwarzschild sphere.

One can imagine a particle having an elementary
magnetic charge, given by the Dirac relation
(\cite{Dirac1948})~~ $\widetilde M=\frac{137}{2}e $
 ~~(with~ $e$~ the electron charge) and a Planckian mass~
 $m=m_{Pl}$. In general relativistic units
one writes
\begin{equation} \label{84}
\widetilde M\approx 9.1\times 10^{-33} {\mathrm{cm}}~,
\qquad
 m\approx 1.608 \times 10^{-33}{\mathrm{cm}}~.
\end{equation}
and this yield
\begin{equation} \label{85}
r_s= 3.216 \times 10^{-33} {\mathrm{cm}}~, \qquad
 r_b=3.53 \times 10^{-1} \mathrm{cm}~.
\end{equation}

From the metric (cf. (\ref{77}), (\ref{78}))
discussed above
\begin{equation} \label{86}
ds^2=\left(1-\frac{2m}{r}(1-K^2 r^2)^{1/2}\right)
dt^2-\frac{1}{\left(1-K^2 r^2\right)
\left[1 -\frac{2m}{r}(1-K^2 r^2)^{1/2}\right]}
dr^2-r^2 d \Omega^2~,
\end{equation}
we can turn to a new metric  with the radial
variable $R$ by:
\begin{equation} \label{87}
r^2=\frac{R^2}{1+K^2R^2}~;~~~~~R^2=
\frac{r^2}{1-K^2r^2}~,
\end{equation}
one readily sees that the ranges are
\begin{equation} \label{88}
0<R<\infty ~,
\end{equation}
while for the old radial variable $r$  we
had (\ref{81}).  For the line-element with
the new variable we get
\begin{equation} \label{89}
ds^2=\left(1-\frac{2m}{R}\right)dt^2-\frac{1}
{(1+K^2R^2)^2 (1-2m/R)}dR^2-
\frac{R^2}{1+K^2R^2} d\Omega~.
\end{equation}
and for the magnetic field strenght
\begin{equation} \label{90}
\widetilde\Phi_{01}= \frac{\widetilde M}{R^2}~.
\end{equation}

The metric (\ref{86}), and field strength
(\ref{79}), or alternativelly (\ref{89}),
and (\ref{90}), may be treated as representing
 a magnetic monopole. It is worth noting that
according to (\ref{75}) a massless electric
monopole may exist. But, unlike it, magnetic
monopoles have to be massive (cf. (\ref{76}) -
 (\ref{79}), and (\ref{80}), (\ref{89})).

\section{Proca Field Accompanying a Magnetic Monopole}

In the previous section a magnetic monopole in
vacuum was considered. We took ~$\kappa=0 $~,
so that massive photons were excluded from the
scenario. But as  mentioned above ( see the
discussion after (\ref{43})) magnetic mopoles
interact just by means of these massive photons.
For that reason it would be useful to consider
the case of vacuum, but with ~$\kappa\neq 0$,
at least  in general.  This situation is
characterized by:
\begin{equation} \label{91}
T^{\mu\nu}=0~,\quad J^{\mu}=0~,
\quad L^{\mu}=0~, \quad Q=0~,
\quad \kappa\neq 0~.
\end{equation}
so that the strength of the magnetic field is
given as previous by equation
(\ref{64}), while from (\ref{56}) - (\ref{58})
 one obtains
 for the stregth of the electric field
\begin{equation} \label{92}
w'=  \kappa^2 ~\frac{e^{1/2(\lambda+\nu)}}
{r^2} I(r)=   \kappa^2 ~\frac{e^{1/2(\lambda+\nu)}}
{r^2}\int_{0}^{r} e^{1/2(\lambda-\nu)} w r^2 dr~.
\end{equation}
With (\ref{91}), and (\ref{92}), we can rewrite
equation (\ref{51}) as
\begin{equation} \label{93}
e^{-\lambda}\left(-\frac{\lambda'}{r}+
\frac{1}{r^2}\right)-\frac{1}{r^2} =-
\kappa^4 \frac{I~^2}{r^4}+\kappa^2
e^{-\nu} w^2 - 3e^{-\nu} V^2  ~~.
\end{equation}
Further we rewrite the difference between
 (\ref{52}), and (\ref{51}) as
\begin{equation} \label{94}
\frac{e^{-\lambda}(\lambda'+\nu~')}{r}=
2e^{-\nu}~[ V^2-\kappa^2 w^2]
\end {equation}
We have also the  energy relation that follows
 by  (\ref{91}) from  (\ref{61})
\begin{equation} \label{95}
2 \kappa^2 w w' + V V' - V^2 \nu~'=0~.
\end{equation}
Thus, as usually, we have two Einstein
equations (\ref{93}), and (\ref{94}), two
"equations of state" for the fields (\ref{64}),
 and(\ref{92}), and the energy condition
(\ref{95}), i. e. a complete system of equations.

One can integrate (\ref{95}), rewriting this as
\begin{equation} \label{96}
\kappa^2 w^2+ \frac{1}{2}~ V^2= 2\int V^2 \nu~' dr
\end{equation}
Recalling that ~$\nu$~ is an increasing function
beyond the Schwarzschild surface we see that the
signs in equation (\ref{96}) are correct.
By (\ref{96})
we can express the $w$-field in terms of the $V$-field.

For a moment let us go back to (\ref{95}).
Introducing an auxiliary function
\begin{equation} \label{97}
z=V^2~.
 \end{equation}
 one  can rewrite (\ref{95}) as
\begin{equation} \label{98}
\frac{1}{2} ~z~' -z~\nu~'=-\kappa^2 (w^2)~'
\end{equation}
and from this one obtains
\begin{equation} \label{99}
V^2 \equiv z= e^{2\nu} \left[ -2 \kappa^2
\int e^{-2\nu} (w^2)~'~ dr +K^2
\right] ~;
\end{equation}
where by comparison with (\ref{68}) the
integration constant is set equal to $K^2$.
We can also rewrite:
\begin{equation} \label{100}
V= e^{\nu} \left[-2\kappa^2 e^{-2\nu} w^2 -4
\kappa^2 \int e^{-2\nu} w^2 \nu~' dr +K^2
\right]^{1/2} ~;
\end{equation}
Differentiating (\ref{100}), comparing with
(\ref{64}), and making use of (\ref{92}) we
 obtain the following relation:
\begin{eqnarray} \label{101}
\left[-2\kappa^2 e^{-2\nu} w^2 -4 \kappa^2
\int e^{-2\nu} w^2 \nu~' dr +K^2 \right]e^{\nu}
r^2 \nu~'-2\kappa^4 e^{1/2(\lambda-\nu)}w\int
e^{1/2(\lambda-\nu)}wr^2dr
\\
\nonumber
-\frac{\widetilde M }{\varepsilon}e^{1/2(
\lambda+\nu)}
\left[-2\kappa^2 e^{-2\nu} w^2 -4 \kappa^2
\int e^{-2\nu} w^2 \nu~' dr +K^2 \right]^{1/2}
= 0 ~;
\end{eqnarray}
Equation (\ref{101}) may be regarded as the
compatibility condition between the energy
 relation
(\ref{95}), and the field equations (\ref{64}),
 and (\ref{92}). Thus one can choose an
appropriate Weyl function $w$, then  from
(\ref{100}) one  obtains $V$, and after
this one can solve (\ref{93}), and (\ref{94}).
However, in order to consider the interaction
between two resting magnetically charged bodies,
one has to turn from spherical symmetry to
cylindrical one.  These procedures will be carried out
in a subsequent paper.
%%%%%555555555555%%%%%%%%%%%%%%%%%%5557666%%%%%%%%%%%%%%%%%%%%%%%%%%%%%%%%%
\section{Discussion}
The Weyl geometry \cite{Weyl1919} is doubtless
 the most aesthetic generalization of the
Riemannian geometry, the last being the framework
 of general relativity.
Dirac \cite{Dirac1973} modified the Weyl theory.
 In order to build up an action integral,
which is coordinate invariant, and  gauge
invariant, and which agrees with the general
relativity theory, Dirac  introduced  a scalar
gauging function,  $\beta $. The   Weyl-Dirac
theory offers a complete basis for deriving
gravitation, and  electromagnetism from geometry
 (cf. e.g. \cite{Isr.Rosen1983}, \cite{Israelit1989}).

In a previous paper \cite{Israelit1996} it
was shown, that extending the Weyl-Dirac theory
  to a framework including torsion, one can
 build up a geometrically based Torsional
Weyl-Dirac Electrodynamics.   This theory,
 possessing intrinsic electric and magnetic
currents and admitting massive photons, is
invariant under Weyl gauge transformations,
and in the Eistein gauge ($\beta=1$) it
takes a simple form.
In the limiting case the theory turns into
the Einstein-Maxwell theory.

In this work we have considered some crucial
problems and investigated physical phenomena
that occur in the discussed framework .
In section 3. the energy-momentum conservation
law (\ref{34}) is treated. It follows that
in absence of fields, induced by magnetic
charges, the photon mass must be set zero
(A similar result, obtained by quite
different methods, may be found in a recent
 paper by R. K\"{u}hne \cite{Kuhne}   ).
Thus in the limiting Einstein-Maxwell case
 the photon are massless in accordance
with classical electrodynamics. From the same
 conservation law we have also derived the
equation of motion of charged test particles
(\ref{42}), and (\ref{43}). It turns out that
the motion of an electrically charged particle
 is affected by the presence of both, electric
sources and magnetic sources. This interaction
may take place either with massless or with
massive photons. On the other hand, on a magnetically
 charged particle the electric sources act in
any case, while the bodies carrying  magnetic
charge  can act only by means of massive photons.
 Thus, massive photons are responsible for the
interaction between two magnetic monopoles.
In sections 4., 5. several spherically symmetric
 solutions are considered. In vacuum a solution
is obtained that gives either the
Reissner-Nordstr\o m solution describing
an electric monopole, or the metric and field
created by a magnetic monopole. An interesting
 consequence is the absence of massless magnetic
 monopoles, while in the framework of the
 Reissner-Nordstr\o m solution we can imagine
 a massless charged particle.

In the theory discussed above magnetic interaction
is transmitted by photons having a finite mass.
The attractive idea of massive photons was
discussed during the last 50 years by many
physicists beginning with de Broglie \cite{de Broglie}
 in 1934. Later, Bondi and Lyttleton \cite{Bondi}
linked massive photons with the cosmological
constant ~$\Lambda$~. In the modern interpretation
one could claim that massive photons cause a
dark matter effect in the universe. There exists
 also a speed-of-light catastrophe in nature,
 which was elegantly displayed by Ardavan
\cite{Ardavan}. Massive photons may help to
 avoid it. Not only in classical  theories,
but also in quantum field theories a zero-mass
 photon leads to difficulties and contradictions.
 Coleman and Weinberg \cite{Coleman} claim that
the photon acquire a mass as a result of radiative
corrections.
Concerning the value of the photons mass, we see
no plausible theoretical tools, which may enable
fixing ~$m_{\gamma}$~ in our framework. Thus one
has to consider experimental estimations.
 In a comprehensive review article by Golhaber,
 and Nieto, that appeared in 1971
\cite{Goldhaber} we find the value $m_{\gamma}
\leq4\times 10^{-48} {\rm{g}}\doteq 2.25\times 10^{-15}
 {\rm{eV}}$, while from a recent Review article
 \cite{Review} we have
 $m_{\gamma}\leq 5.34 \times 10^{-60}
 {\rm{g}}\doteq 3\times 10^{-27}  {\rm{eV}} $.
 If we adopt the latter  value we get from
(\ref{39})~~ for our parameter ~$\kappa\leq 1.53\times
 10^{-22}
 ( {\rm cm^{-1}})$.

There are very few experimental datae
concerning properties of magnetic monopoles.
 If one believes that they were produced in the
 very early universe during the first ~~$10^{-34}
 {\rm{seconds}}$~ \cite{Cabrera}, their mass is
$  m\approx 10^{19} {\rm{MeV}}\doteq 1.8 \times
10^{-8} {\rm {g}} $.
In interesting experiments during the last decade
 (cf. e.g. \cite{Barish},
\cite{Bertani}, \cite{Orito} )  monopoles having mass
~~$10^{6} {\rm{MeV}}\leq m \leq 10^{18}
{\rm{MeV}}$~~
 were recorded. One can hope that heavier monopoles
will be recorded in the future.

If one would like to have a "standard" magnetic
 monopole,
the Planckian mass
~$m_{Pl}= 1.22\times10^{22}{\rm {MeV}}$
 would look very attractive. One could also
imagine
 an elementary carrier of magnetic charge,
as a non-singular
  entity, having charge, mass, and dimensions,
 and filled with matter possessing mass density,
as well magnetic charge density.
Building up a model of this particle would be an
interesting challenge.

%%%%%%%%%%%%%%%%%%%%%%%%%%%%%%%%%%%%%%%%%%%%%%%%%%%%%%%%%%%%%%%%%%%%%

\section*{Acknowledgments}
{The author takes this opportunity to express
his cordial thanks to Professor HEINZ DEHNEN
for very interesting discussions.}

%%%%%%%%%%%%%%%%%%%%%%%%%%%%%%%%%%%%%%%%%%%%%%%%%%%%%%%%%%%%%%%%%%%%%

\end{document}